\documentclass[aps,twocolumn,nofootinbib,superscriptaddress,prl,showpacs]{revtex4-1}
\usepackage{amsmath}
\usepackage{epsfig}
\def\bea#1\eea{\begin{align}#1\end{align}} 
\newcommand{\nnu}{\nonumber\\}
\newcommand{\bef}{\begin{figure}[htb]\centering}
\newcommand{\eef}{\end{figure}}

\begin{document}
\title{Next-to-Leading QCD Factorization for Semi-Inclusive Deep Inelastic \\ Scattering at Twist-4}

\date{\today}

\author{Zhong-Bo Kang}
\affiliation{Theoretical Division, 
                   Los Alamos National Laboratory, 
                   Los Alamos, New Mexico 87545, USA}

\author{Enke Wang}
\affiliation{Institute of Particle Physics and Key Laboratory of Lepton and Quark Physics (MOE), 
                   Central China Normal University, 
                   Wuhan 430079, China}

\author{Xin-Nian Wang}
\affiliation{Institute of Particle Physics and Key Laboratory of Lepton and Quark Physics (MOE),
                  Central China Normal University, 
                  Wuhan 430079, China}
\affiliation{Nuclear Science Division, 
                   Lawrence Berkeley National Laboratory, 
                   Berkeley, California 94720, USA}

\author{Hongxi Xing}
\affiliation{Theoretical Division, 
                   Los Alamos National Laboratory, 
                   Los Alamos, New Mexico 87545, USA}
\affiliation{Institute of Particle Physics and Key Laboratory of Lepton and Quark Physics (MOE), 
                   Central China Normal University, 
                   Wuhan 430079, China}
\affiliation{Interdisciplinary Center for Theoretical Study and Department of Modern Physics, 
                   University of Science and Technology of China, 
                   Hefei 230026, China}

\begin{abstract}
Within the framework of a high-twist approach, we calculate the next-to-leading order (NLO) perturbative QCD corrections to the transverse momentum broadening in semi-inclusive hadron production in deeply inelastic $e+A$ collisions, as well as lepton pair production in $p+A$ collisions. With explicit calculations of both real and virtual contributions, we verify, for the first time, the factorization theorem at twist 4 in NLO for the nuclear-enhanced transverse momentum weighted differential cross section and demonstrate the universality of the associated twist-4 quark-gluon correlation function. We also identify the QCD evolution equation for the twist-4 quark-gluon correlation function in a large nucleus, which can be solved to determine the scale dependence of the jet transport parameter in the study of jet quenching.
\end{abstract}


\maketitle


Multiple scatterings of  energetic partons inside cold or hot nuclear matter play an important role in the study of the QCD medium in high-energy lepton-nucleus, hadron-nucleus and nucleus-nucleus collisions. They lead to parton energy loss and transverse momentum 
broadening \cite{Gyulassy:1993hr, BDMPS, Gyulassy:2000er,Zakharov:1997uu} that are responsible for the observed jet quenching phenomena in semi-inclusive deeply inelastic scattering (SIDIS)  \cite{Wang:2009qb} and high-energy heavy-ion collisions \cite{Majumder:2010qh}. Though there has been significant progress in the study of parton energy loss \cite{Gyulassy:2003mc,Kovner:2003zj}, radiative correction to transverse momentum broadening \cite{Mueller:2012bn}, and efforts to include the effect of multiple gluon emission \cite{MehtarTani:2010ma,Blaizot:2013hx, Fickinger:2013xwa}, the main theoretical uncertainty in current jet quenching studies arises from the logarithmic dependence on the kinematic cut-off in the leading-order (LO) calculation of parton energy loss \cite{Armesto:2011ht} and the lack of a proof of factorization of hard scattering and the medium properties. A complete  next-to-leading order (NLO) calculation of parton energy loss and an analysis of factorization at NLO are essential for future quantitative understanding of ever more precise data on jet quenching from high-energy SIDIS and heavy-ion collision experiments.

One of the approaches to parton energy loss \cite{Wang:2001ifa,Xing:2011fb} and transverse momentum broadening \cite{Guo:1998rd,Guo:2000eu,Fries:2002mu,Majumder:2007hx,Liang:2008vz,Kang:2008us,Schafer:2013mza} is based on high-twist formalism that assumes collinear factorization \cite{Luo:1992fz,Qiu:1990xy,Collins:1989gx}. Within such an approach, one carries out collinear expansion of hard parts of multiple scattering amplitudes and reorganizes the final results in terms of power corrections. Dominant contributions often depend on high-twist matrix elements of the nuclear state that are enhanced by the nuclear size. So far most studies have focused on double parton scattering and proofs of factorization are only limited to LO analyses \cite{Qiu:1990xy}.

In this Letter, we will carry out, for the first time, a complete NLO analysis of the twist-4 contributions to the transverse momentum weighted cross section of SIDIS. In particular, we consider contributions of quark rescattering with partons from another nucleon inside the nucleus. Such contributions are proportional to the nuclear size $A^{1/3}$. For large nucleus $A\gg 1$, we neglect other $A$-independent higher-twist contributions,  for example,  any twist-4 fragmentation correlation contributions that have no $A^{1/3}$ enhancement \cite{Braun:2008ia}. 
We will calculate explicitly the real and virtual corrections up to one-loop order to the twist-4 contributions. We verify the factorization  theorem at twist 4 in NLO by demonstrating the cancellation of infrared divergences and renormalization of the twist-4 parton correlation functions. Our results not only provide a complete NLO calculation of the transverse momentum weighted cross section and, therefore, transverse momentum broadening at twist 4, but also pave the way to the proof of QCD factorization for higher-twist hard processes and complete NLO calculation of jet quenching in medium. 

In SIDIS of the hadron production off a large nucleus,
\bea
e(L_1)+A(p)\rightarrow e(L_2)+h(\ell_h)+X,
\eea
we consider the invariant mass  of the virtual photon $Q^2 = -q^2=-(L_2 - L_1)^2$ is large, where $p$ is the average momentum per nucleon in the nucleus with the atomic number $A$ and $\ell_h$ is the momentum of a final-state hadron $h$. Higher-twist contributions to the cross section from multiple scatterings are normally suppressed by powers of $1/Q^2$. The coefficients, which are determined by high-twist nuclear matrix elements, are, however, enhanced by the nuclear size $A^{1/3}$ \cite{Luo:1992fz}. The transverse momentum 
broadening,  $\Delta\langle\ell_{hT}^2\rangle=\langle\ell_{hT}^2\rangle_{eA}-\langle\ell_{hT}^2\rangle_{ep}$,
is defined as the difference between the average squared transverse momentum of the observed hadron produced on a nuclear target and 
that on a proton target. Leading contributions in the high-twist approach come from double scatterings,
\bea
\Delta \langle\ell_{hT}^2\rangle \approx 
\frac{d\langle \ell_{hT}^2\sigma^D\rangle}{d{\cal PS}}  
\left/
\frac{d\sigma}{d{\cal PS}}\right.,
\label{master}
\eea
where the phase space $d{\cal PS} = dx_Bdydz_h$ with $x_B=Q^2/(2p\cdot q)$, $y=p\cdot q/p\cdot L_1$ and $z_h=p\cdot \ell_h/p\cdot q$ the  usual SIDIS variables \cite{Kang:2012ns}.
The transverse momentum $\ell_{hT}^2$-weighted cross section from double scattering
\bea
\frac{d\langle \ell_{hT}^2\sigma^D\rangle}{d{\cal PS}} \equiv \int d\ell_{hT}^2 \ell_{hT}^2 \frac{d\sigma^D}{d{\cal PS}d\ell_{hT}^2},
\eea
will be the focus of our NLO analysis in this Letter.

\bef
\psfig{file=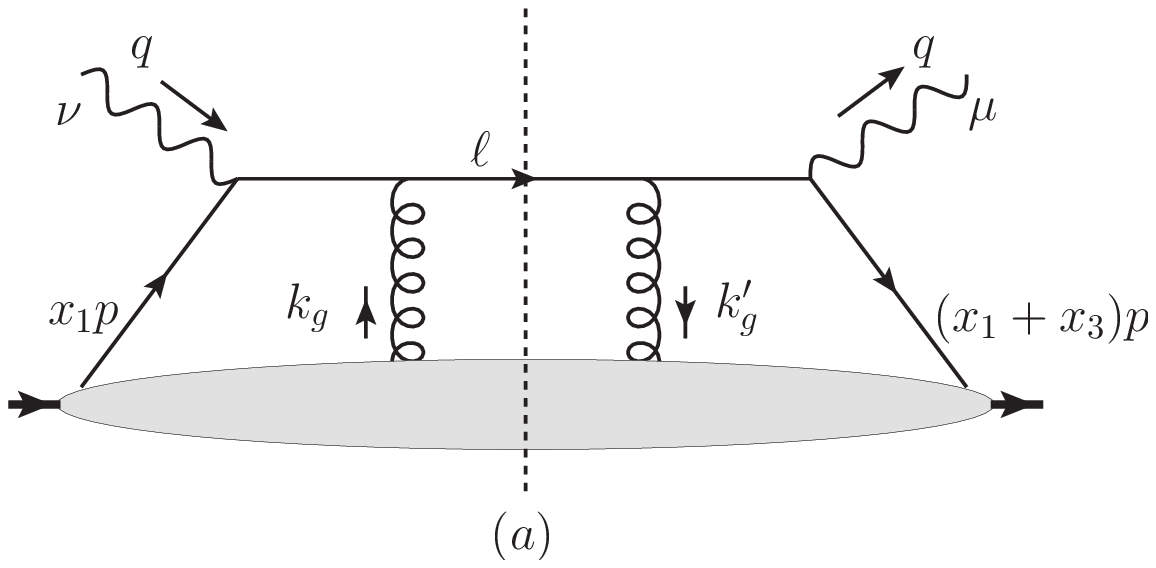, width=0.55\columnwidth}
\hskip 0.1in
\psfig{file=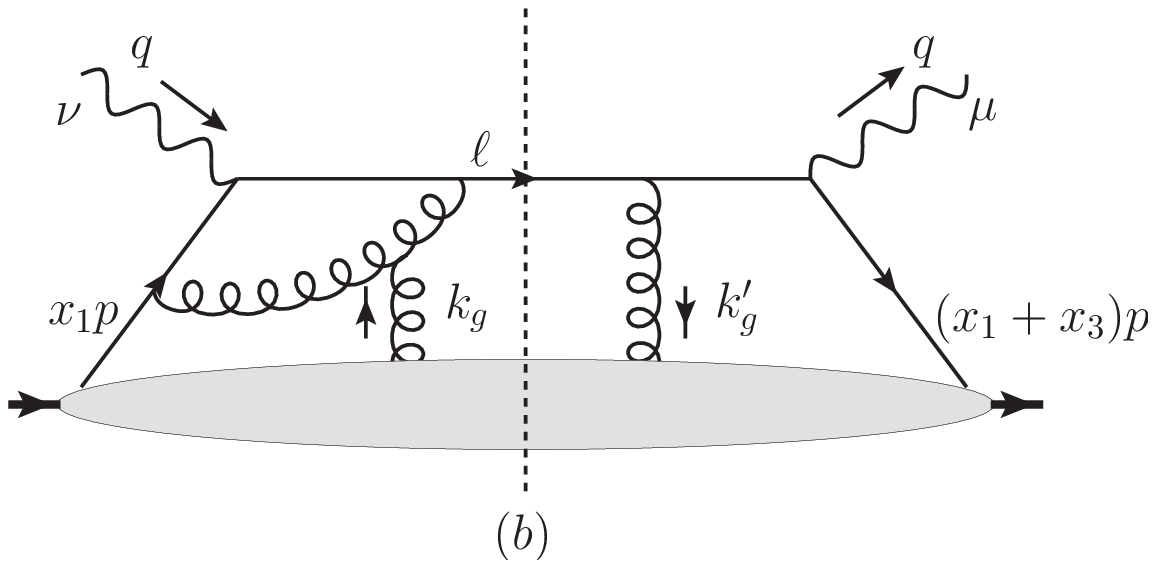, width=0.6\columnwidth}
\hskip 0.1in
\psfig{file=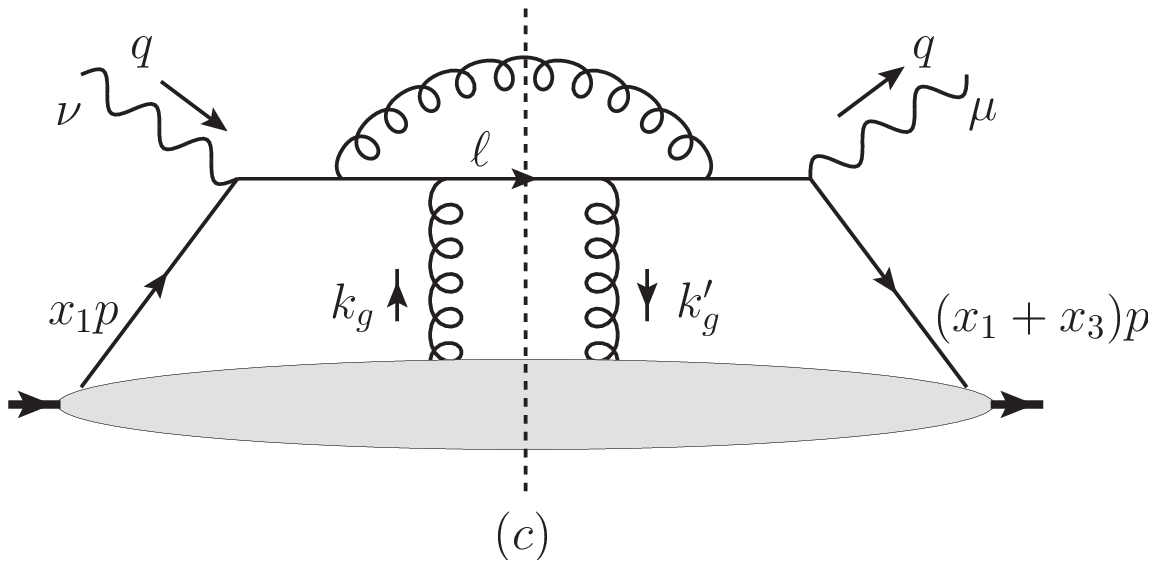, width=0.6\columnwidth}
\caption{Sample Feynman diagrams for double scattering contributions to the $\ell_{hT}^2$-weighted differential cross section from (a) leading-order  (b) NLO virtual, and (c)  NLO real processes. The gluon momenta involved in the double scattering are $k_g=x_2 p+k_T$ and $k_g'=(x_2-x_3)p+k_T$.}
\label{diagram}
\eef

Double scattering contributions manifest themselves as twist-4 power corrections to the differential cross section. To derive these contributions, a generalized factorization theorem, the so-called high-twist power expansion approach, was developed some time ago \cite{Luo:1992fz}. It involves the so-called collinear expansion, i.e., expand the hard scattering contribution around the vanishing parton transverse momentum as in Fig.~\ref{diagram}. At leading order, the contribution is given by Fig.~\ref{diagram}(a). Within such a high-twist approach, a straightforward calculation leads to a simple LO result,
\bea
\frac{d\langle \ell_{hT}^2\sigma^D\rangle}{d{\cal PS}} & = \sigma_h \sum_q e_q^2
\int \frac{dz}{z} D_{h/q}(z) \int  \frac{dx}{x} T_{qg}(x,0,0) 
\nnu
& \times  \delta(1-{\hat x})\delta(1-{\hat z}),
\label{eq-LO}
\eea
where $\hat x= x_B/x$, $\hat z=p\cdot \ell/p\cdot q = z_h/z$ with $\ell$ the momentum of the final state quark, and $\sigma_h = (4\pi^2\alpha_{\rm s}z_h^2/N_c) \sigma_0$ with $\sigma_0$ given by
\bea
\sigma_0= \frac{2\pi\alpha_{\rm em}^2}{Q^2}\frac{1+(1-y)^2}{y} (1-\epsilon),
\eea
where $\epsilon$ is introduced since $n=4-2\epsilon$ dimensions will be used for dimensional regularization in NLO calculations. The twist-4 quark-gluon correlation function is defined as \cite{Wang:2001ifa,Kang:2008us,Osborne:2002st},
\bea
T_{qg}(x_1, x_2, x_3)
&= \int \frac{dy^-}{2\pi} e^{ix_1p^+y^-}  \int \frac{dy_1^-dy_2^-}{4\pi} 
\nnu
&\hspace{-24pt}\times e^{ix_2p^+(y_1^- - y_2^-)}
 e^{ix_3p^+y_2^-} \theta(y_2^-)\theta(y_1^- - y^-)
\nnu
&\hspace{-24pt}\times  \langle A|{\bar\psi}_q(0) \gamma^+ F_{\sigma}^+(y_2^-)F^{\sigma +}(y_1^-)\psi_q(y^-)|A\rangle,
\eea
which contains the fundamental properties of the nuclear medium as probed by a propagating quark and can be extracted from experimental data, such as the transverse momentum broadening in DIS and DY processes  \cite{Guo:2000eu,Guo:1998rd}.

For the purpose of a NLO analysis of the $\ell_{hT}^2$-weighted differential cross section $d\langle \ell_{hT}^2\sigma^D\rangle/d{\cal PS}$, one needs to evaluate both virtual and radiative corrections to the LO process.  Some example diagrams are shown in Figs.~\ref{diagram}(b) and \ref{diagram}(c) for virtual and real corrections, respectively. Calculations of contributions from these diagrams involve a significant amount of tensor reduction and integration. However, the final results have some simple structures that contain both infrared and collinear divergences.

The final NLO virtual correction to the $\ell_{hT}^2$-weighted cross section at twist 4 is given by the LO result as in Eq.~\eqref{eq-LO}, multiplying by the following factor:
\bea
\frac{\alpha_s}{2\pi} C_F \left(\frac{4\pi\mu^2}{Q^2}\right)^{\epsilon}
\frac{1}{\Gamma(1-\epsilon)} \left[-\frac{2}{\epsilon^2}-\frac{3}{\epsilon}-8\right],
\label{virtual}
\eea
where $\mu$ comes from the replacement $g\to g\mu^{\epsilon}$ in $n=4-2\epsilon$ dimensions. A similar structure also appears in the virtual correction to the transverse momentum weighted spin-dependent cross section at twist 3 \cite{Kang:2012ns,Vogelsang:2009pj}. 

To calculate radiative corrections to the $\ell_{hT}^2$-weighted cross section, we follow established techniques for contour integration and collinear expansion \cite{Wang:2001ifa,Kang:2013ufa}.  We take the central-cut diagram in Fig.~\ref{diagram}(c) as an example, in which the quark radiates one gluon between its interaction with the virtual photon and one additional scattering inside the nucleus. The quark and radiated gluon will eventually fragment into final observed hadrons. 

The on-shell condition for the radiated gluon leads to a $\delta$ function $ \delta(x_1+x_2 - x-x_c)$ and, thus, fixes $x_1 = x+x_c - x_2$, with $x=(Q^2+2q\cdot \ell)/2p\cdot (q-\ell)$ and $x_c =(k_T^2 - 2 k_T\cdot \ell)/2p\cdot (q-\ell)$.
The remaining two momentum fractions $x_2$ and $x_3$ are eliminated via contour integrations which set one of the two quark propagators on each side of the cut line on their mass shells.  When the quark propagator immediately after the quark-photon interaction is on its mass shell, the gluon radiation is induced by the secondary quark-nucleus scattering where the exchanged gluon momentum ($k_g$ or $k_g'$) remains finite in the collinear limit $k_T\to 0$. Thus, this is referred to as a hard scattering. On the other hand, if the quark propagator close to the quark-photon vertex remains off shell, it becomes on shell after radiating a gluon. Since the exchanged gluon momentum becomes zero when $k_T\to 0$ in this case, it is called a soft scattering. The final result of this cut diagram contains soft, hard contributions and their interferences, often referred to as soft-soft, hard-hard, soft-hard, and hard-soft contributions. Besides similar central cut diagrams shown in Fig.~\ref{diagram}(c), we also need to include asymmetrical-cut diagrams which represent interferences between single and triple scattering.

There are, in total, 57 diagrams for soft-soft, hard-hard, and the interferences between soft and hard rescatterings in quark-gluon double scatterings. All these processes contain both soft-collinear and collinear divergences, which can be identified in dimensional regularization as double-pole $1/\epsilon^2$ and single-pole $1/\epsilon$ type divergences, respectively. In addition to the photon+quark channel, we also consider the photon+gluon channel in NLO processes at twist 4 that has only soft-soft contributions and contains only the $1/\epsilon$ type collinear divergences. It involves the gluon-gluon correlation function \cite{Kang:2013ufa},
\bea
T_{gg}(x_1,x_2,x_3) = & \frac{1}{x p^+} \int \frac{dy^{-}}{2\pi}\, e^{ix_1 p^{+}y^{-}}
 \int \frac{dy_1^{-}dy_{2}^{-}}{2\pi} \nnu
&\hspace{-44pt} \times   e^{ix_2p^+(y_1^- - y_2^-)}e^{ix_3p^+y_2^-} \theta(y_{2}^{-})\,  \theta(y_1^{-}-y^{-}) \nnu
& \hspace{-44pt} \times  \langle A| F_\alpha^{~+}(0)
F^{\sigma+}(y_2^-)F^+_{~\sigma}(y_1^-)F^{+\alpha}(y^-)|A\rangle\, .
\label{Tgg}
\eea

Combining real and virtual corrections, we find that the double-pole $1/\epsilon^2$ terms cancel out as required by QCD factorization, and the final result with only the $1/\epsilon$ terms from collinear divergences can be written as 
\bea
\frac{d\langle \ell_{hT}^2\sigma^D\rangle}{d{\cal PS}} & =
\sigma_h \frac{\alpha_s}{2\pi}
\sum_q e_q^2 \int\frac{dz}{z} D_{h/q}(z)\int\frac{dx}{x} \left(-\frac{1}{\hat\epsilon} \right. 
\nnu
& \hspace{-34pt} +\left. \ln\frac{Q^2}{\mu_f^2} \right)
\Big[\delta(1-\hat x) P_{qq}(\hat z) T_{qg}(x,0,0) + \delta(1-\hat z)
\nnu
& \hspace{-34pt} \times \big({\mathcal P}_{qg\to qg} \otimes T_{qg} 
+ P_{qg}(\hat x)T_{gg}(x,0,0) \big)\Big] + \cdots,
\label{Eq-totdiv}
\eea
where $1/\hat\epsilon=1/\epsilon-\gamma_E+\ln(4\pi\mu^2/\mu_f^2)$, and $\mu_f$ is the factorization scale
in the $\overline{\rm MS}$ subtraction scheme that we use.
Here, and in the rest of this Letter, we only write explicitly the divergent pieces, and the finite contributions denoted by ``$\cdots$'' will be published elsewhere \cite{future}.
$P_{qq}(\hat x)$ and $P_{qg}(\hat x)$ are the usual quark-to-quark and gluon-to-quark splitting kernels in the leading-twist DGLAP evolution equations, and the term ${\mathcal P}_{qg\to qg} \otimes T_{qg}$ is defined as
\bea
& \int \frac{dx}{x} {\mathcal P}_{qg\to qg} \otimes T_{qg}  \equiv 
\int \frac{dx}{x} P_{qq}(\hat x) T_{qg}(x, 0, 0) 
\nnu
&+ \frac{C_A}{2} \bigg[ \frac{4}{(1-\hat x)_+} 
T_{qg}(x_B, x-x_B, 0) - \frac{1+\hat x}{(1-\hat x)_+}
\nnu
&\hspace{-2pt} \times \big(T_{qg}(x,0,x_B-x)+T_{qg}(x_B,x-x_B,x-x_B)\big)\bigg].
\eea
According to the analysis of induced gluon spectra in Refs.~\cite{Wang:2001ifa,Xing:2011fb}, the interference between soft and hard contributions corresponds to Landau-Pomeranchuk-Migdal interference which suppresses gluon radiation with large  formation time $t_f=1/(x-x_B)p^+>R_A$. In this region the value of the above integrand is, indeed, suppressed by $\sim x-x_B < 1/(p^+R_A)$.

Following the strategy of collinear factorization procedure, we isolate and absorb the collinear divergences into the renormalization of corresponding nonperturbative functions. The first [$\propto \delta(1-\hat x)$]  collinear divergence from final-state gluon radiation in Eq.~\eqref{Eq-totdiv} 
 can be absorbed into the renormalized quark fragmentation function which satisfies the usual DGLAP evolution equations \cite{Altarelli:1979ub}. 
 The second [$\propto \delta(1-\hat z)$] collinear divergence from initial-state radiation can be absorbed into the renormalized quark-gluon correlation function
\bea
T_{qg}(x_B, 0, 0, \mu_f^2) & =  T_{qg}(x_B,0,0) -\frac{\alpha_s}{2\pi}\frac{1}{\hat\epsilon}\int_{x_B}^1\frac{dx}{x}
\nnu
&\hspace{-34pt} \times \bigg[{\mathcal P}_{qg\to qg} \otimes T_{qg}+P_{qg}(\hat x)T_{gg}(x,0,0)\bigg].
\eea 
The above leads to a new QCD evolution equation for the quark-gluon correlation function
\bea
\frac{\partial}{\partial \ln\mu_f^2} T_{qg}(x_B,0,0,\mu_f^2) & =  \frac{\alpha_s}{2\pi} 
\int_{x_B}^1 \frac{dx}{x} \bigg[{\mathcal P}_{qg\to qg} \otimes T_{qg} 
\nnu
&
+ P_{qg}(\hat x) T_{gg}(x, 0, 0, \mu_f^2)\bigg].
\label{evolution}
\eea
Eq.~\eqref{evolution}, as it stands, is not closed. It is a common feature for higher-twist parton distributions~\cite{Kang:2012ns,Vogelsang:2009pj}. Under certain approximations for the functional form in $x_{i=1-3}$ of the two-parton correlation function, one could obtain a solution to the above evolution equation \cite{future}. In principle, one could also derive an evolution equation for the gluon-gluon correlation function $T_{gg}$, which will be pursued in the future. 

In terms of the renormalized quark fragmentation function and twist-4 quark-gluon correlation function, we can express 
the NLO correction to the $\ell_{hT}^2$-weighted differential cross section as
\bea
\frac{d\langle \ell_{hT}^2\sigma^D\rangle}{d{\cal PS}} & \stackrel{\rm NLO}{=}
\sigma_h\frac{\alpha_s}{2\pi} \sum_q e_q^2 \int_{z_h}^1 \frac{dz}{z}D_{h/q}(z, \mu_f^2) 
\nnu
&\hspace{-24pt} \times
\int_{x_B}^1\frac{dx}{x} \bigg\{\ln\left(\frac{Q^2}{\mu_f^2}\right)  \Big[\delta(1-\hat x)P_{qq}(\hat z) 
\nnu
&
\hspace{-24pt} \times   T_{qg}(x, 0, 0, \mu_f^2)
+ \delta(1-\hat z)
\nnu
&
\hspace{-24pt} \times
\big({\mathcal P}_{qg\to qg}\otimes T_{qg} 
+ P_{qg}(\hat x) T_{gg}(x, 0, 0, \mu_f^2)\big) \Big]
\nnu
&\hspace{-24pt} + H_q^{\rm NLO}(\hat x,\hat z) \otimes T_{qg} + H_g^{\rm NLO}(\hat x,\hat z) \otimes T_{gg} \bigg\},
\eea
where the last line in the equation comes from the finite contribution from quark-gluon and gluon-gluon double scatterings after subtraction of collinear divergences in the $\overline{\rm MS}$ scheme, and will be presented elsewhere \cite{future}. So far, our results verify for the first time the factorization of $\ell_{hT}^2$-weighted differential cross section at twist 4 in NLO. The collinear divergences associated with the quark fragmentation function and twist-4 quark-gluon correlation function are factorized, and one is left only with finite hard coefficient functions, which also depend on the factorization scale.  One should also consider contributions from double quark scattering \cite{Schafer:2007xh} and hadron production from gluon fragmentation for more complete NLO calculations \cite{future}.

We have also verified the factorization for the transverse momentum weighted differential cross section of Drell-Yan lepton pair production in p+A collisions at twist 4 in NLO \cite{future}. It  contains the same twist-4 quark-gluon correlation function which follows the same evolution equation as in SIDIS in Eq.~\eqref{evolution}. This confirms, for the first time, the collinear factorization for twist-4 observables at the NLO, and demonstrates the universality of the associated twist-4 correlation functions.  Therefore, the properties of nuclear matter contained in the universal correlation functions as probed by a propagating parton are independent of the hard processes that create the fast partons. 

Under the approximation of a large and loosely bound nucleus, in which momentum and spatial correlations 
of two nucleons can be neglected \cite{Osborne:2002st,Kang:2013ufa},  one can express $T_{qg}(x_B,0,0,\mu_f^2)$ in a factorized 
form  \cite{CasalderreySolana:2007sw},
\begin{equation}
T_{qg}(x_B,0,0,\mu_f^2) \approx \frac{N_c}{4\pi^2\alpha_{\rm s}} f_{q/A}(x_B, \mu_f^2) \int dy^- \hat {q}(\mu_f^2,y^-),
\end{equation}
in terms of the quark distribution function $f_{q/A}(x_B, \mu_f^2)$ and the jet transport parameter $\hat {q}(\mu_f^2,y^-)$.
The new evolution equation in Eq.~(\ref{evolution}) can then determine the scale dependence of $\hat {q}(\mu_f^2,y^-)$. Such a scale dependence will have important consequences on quantitative studies of jet quenching at NLO.

In summary,  we have demonstrated, for the first time, the factorization of the nuclear-enhanced twist-4 contributions to the transverse momentum weighted differential cross section of  SIDIS as well as Drell-Yan lepton pair production in $p+A$ collisions in NLO within the high-twist formalism. With explicit calculations, we have shown that the soft divergences cancel out between real and virtual diagrams, while the collinear divergences can be absorbed into the renormalized fragmentation (or parton distributions) functions, as well as the twist-4 quark-gluon correlation function. Our results demonstrate, for the first time, the universality of such a twist-4 quark-gluon correlation function which contains the properties of nuclear matter such as the jet transport parameter as probed by a propagating parton in different processes. We also derived a QCD evolution equation for the twist-4 quark-gluon correlation function. This will provide a framework to determine the QCD scale dependence of the jet transport parameter. These complete NLO corrections together with the scale dependence of the twist-4 correlation functions, will provide more quantitative  descriptions of the transverse momentum broadening due to multiple parton scattering in nuclear medium, and pave the way for the NLO calculation of jet quenching. 

We thank J.~W. Qiu and I. Vitev for helpful discussions and Y.-Q. Ma for his MATHEMATICA package to calculate Feynman diagrams. This work is supported by U.S. DOE under Contract No.~DE-AC52-06NA25396 and No. DE-AC02-05CH11231, and within the framework of the JET Collaboration, the Major State Basic Research Development Program in China (No. 2014CB845404), and the NSFC under Grants No.~11221504 and No. 10825523, and China MOST under Grant No. 2014DFG02050. 


\end{document}